\documentclass[11pt,twoside]{article}
\usepackage{asp2006}
\usepackage{epsf}
\usepackage{graphicx}
\usepackage{lscape}
\def\edcomment#1{\iffalse\marginpar{\raggedright\sl#1\/}\else\relax\fi}
\reversemarginpar

\begin{document}
\title{Sunspot light-bridges - a bridge between the photosphere and the corona?}
\author{Sarah Matthews}
\affil{UCL Mullard Space Science Laboratory, Holmbury St. Mary, Dorking, Surrey RH5 6NT, U.K.}
\author{Deborah Baker and Santiago Vargas Dom\'inguez}
\affil{UCL Mullard Space Science Laboratory, Holmbury St. Mary, Dorking, Surrey RH5 6NT, U.K.}

\begin{abstract}
\noindent Recent observations of sunspot light-bridges have shed new light on the fact that they are often 
associated with significant chromospheric activity. In particular chromospheric jets \citep{shimizu2009}
persisting over a period of days have been identifies, sometimes associated with large downflows at the 
photospheric level \citep{louis2009}. One possible explanation for this activity is reconnection low 
in the atmosphere. Light-bridges have also been associated with a constant brightness enhancement in the 
1600 \AA\ passband of TRACE, and the heating of 1 MK loops. Using data from EIS, SOT and STEREO EUVI we 
investigate the response of the transition region and lower corona to the presence of a light-bridge and specific 
periods of chromospheric activity.
\end{abstract}
\vspace{-0.5cm}
\section{Introduction}
Sunspot light-bridges are bright lanes of material that divide the umbra. Their appearance signifies 
the reestablishment of the granulation within the spot, and often indicates the beginning of fragmentation 
of the spot itself \citep{vazquez1973}. Observations 
from the Swedish Solar Telescope \citep{scharmer2003} by \cite{berger2003} showed for the first 
time that dark central lanes are common features of strong light-bridges. The magnetic field within 
light-bridges has been observed to be both weaker and more inclined than 
in the surrounding umbra \citep[e.g.][]{leka1997} and their increased brightness 
relative to the surrounding umbra is a clear indication that the plasma temperature 
in this region is higher. It has also been noted that light-bridges often show enhanced chromospheric 
activity, with H$\alpha$ surges and chomospheric jets reported in a number of cases \citep{roy1973,asai2001,bharti2007,shimizu2009}.
 \cite{berger2003} also found a constant brightness enhancement above a light-bridge in TRACE 1600 \AA\ observations, while
 \cite{katsukawa2007} found that light-bridge  formation was spatially and temporally coincident with the heating of $\approx$ 1 MK loops as observed 
by TRACE.  Light-bridges thus seem important for releasing magnetic energy stored in the spot as 
well as in its decay. In this work we investigate the extent to which the presence of a sunspot 
light-bridge affects the overlying transition region and corona.

\section{Observations}

AR 10953 was observed by Hinode during the period 28 April 2007 
to 3 May 2007 \citep[e.g.][]{shimizu2009}, and by STEREO EUVI 
\citep{howard2008}. SOT observed with the broad-band filter imager (BFI) and 
in the G-band, H$\alpha$ and Ca II H passbands. The spectropolarimter (SP) performed 
fast maps of the region obtaining the full-polarization 
states of the Fe lines at 630.15 and 630.25 nm at 0.32$^{\prime\prime}$ resolution. 
EIS \citep{culhane2007} observed between 28 Apr 07 to 2 May 07, first with the 
266$^{\prime\prime}$ slot in the Fe XV 284 \AA\ line with 20s cadence. The Fe XV line is 
free from blends over the slot dispersion, and thus represents monochromatic emission at 
approximately 1.6 MK. 
During the remainder of the observing period, EIS produced rasters with both the 1$^{\prime\prime}$ 
and 2$^{\prime\prime}$ slits. Here we focus on the He II 256.32 \AA\, Si VII 275.35 \AA\, 
Fe XII 195.12 \AA\ and Fe XV 284.16 \AA\ lines that provide broad 
temperature coverage from the chromosphere to the corona. Standard calibration procedures were 
applied to the EIS data and cross-correlation techniques used to remove the orbital variation and 
spacecraft jitter that affects the 266 $^{\prime\prime}$ slot images. Line-of-sight velocity maps were 
derived from the EIS raster data using a single Gaussian fit (\emph{mpfit}). He II 256.32 is substantially 
blended with Fe XIII  256.42, Fe XII 256.41 and Si X 256.37, but for disk observations of an active region such as 
this the He II contribution should contribute 80\% or more to the blend \citep{young2007}.  
Data from the SP were corrected for dark current, flat-field and cosmic rays with standard solarsoft 
routines and then inverted using the full atmosphere inversion code LILIA (Socas Navarro, 2001). The 
intrinsic 180-degree azimuth ambiguity was resolved using the Non-Potential magnetic 
Field Calculation (NPFC) method by \cite{georgoulis2005}. Standard calibration procedures were applied to 
the EUVI data and co-alignment between SOT, EIS and EUVI was achieved through first co-aligning SOT  
H$\alpha$ images with EIS He II 256 \AA\ and EUVI He II 304 \AA\ .

\begin{figure}[!ht]
\includegraphics[angle=-90,width=.95\linewidth]{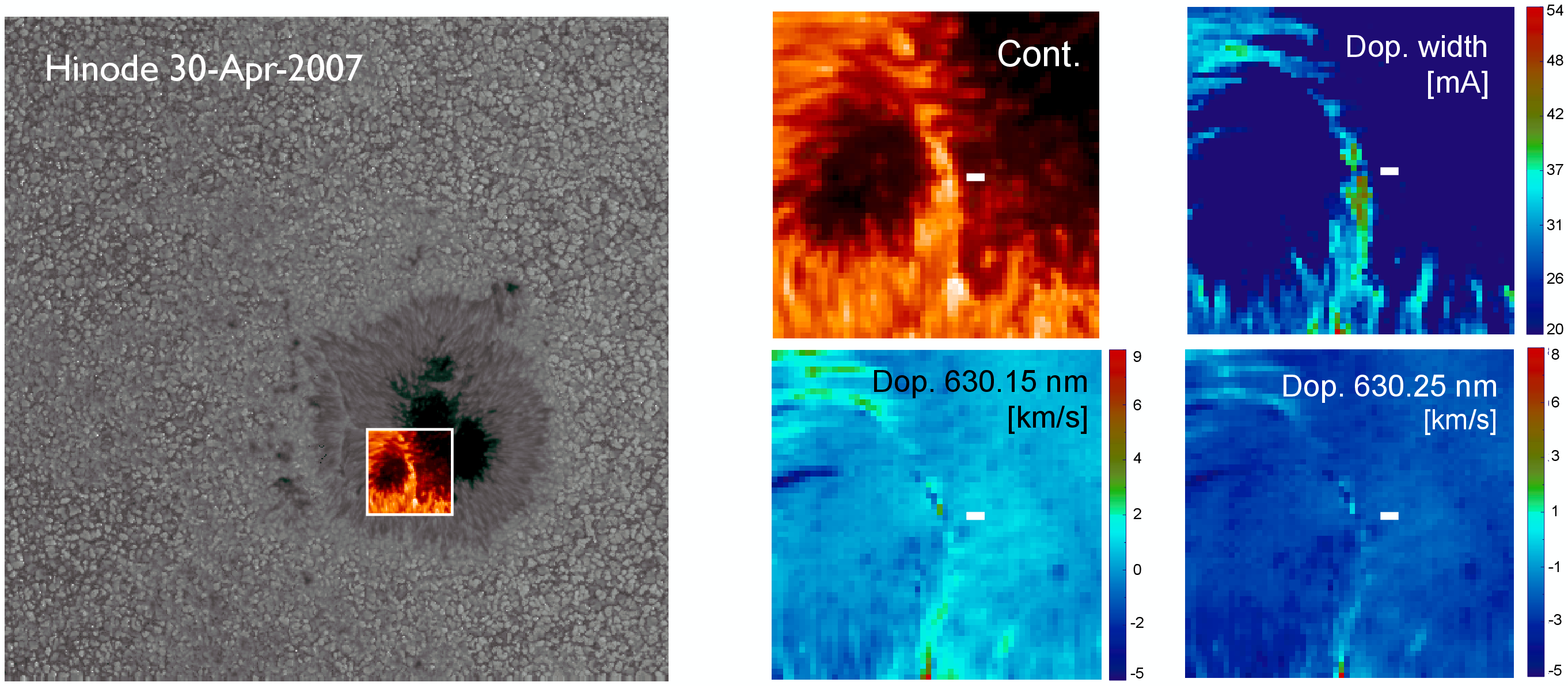}
\caption{Left: SOT SP continuum image of the sunspot on 30 April 07; right: SOT close
up of the light-bridge showing continuum intensity, Doppler width and Doppler velocity derived from
the LILIA inversion code.}
\end{figure}

\section{Connections between the photosphere, chromosphere and corona}

We concentrate on periods of good EIS data coverage to study the 
coronal counterpart 
(if any) of the light-bridge, and thus focus here on the 30 April 2007 during 
which EIS was performing 
raster scans of the region. SOT SP performed several scans during 
this period. Figure 1 shows an SOT continuum image of the sunspot and 
light-bridge (left) and a close up of the light-bridge showing photospheric Doppler velocities
within the structure. Figure 2 shows the coronal emission within the active region and its
relationship to the chromospheric Ca II emission. Light-curves of the intensity within the 
light-bridge and its immediate vicinity form Ca II and EUVI 171 data show no clear relationship 
between variations in the corona and the chromosphere.

\begin{figure}[!ht]
\hspace{-2cm}\includegraphics[angle=-90,width=1.3\linewidth]{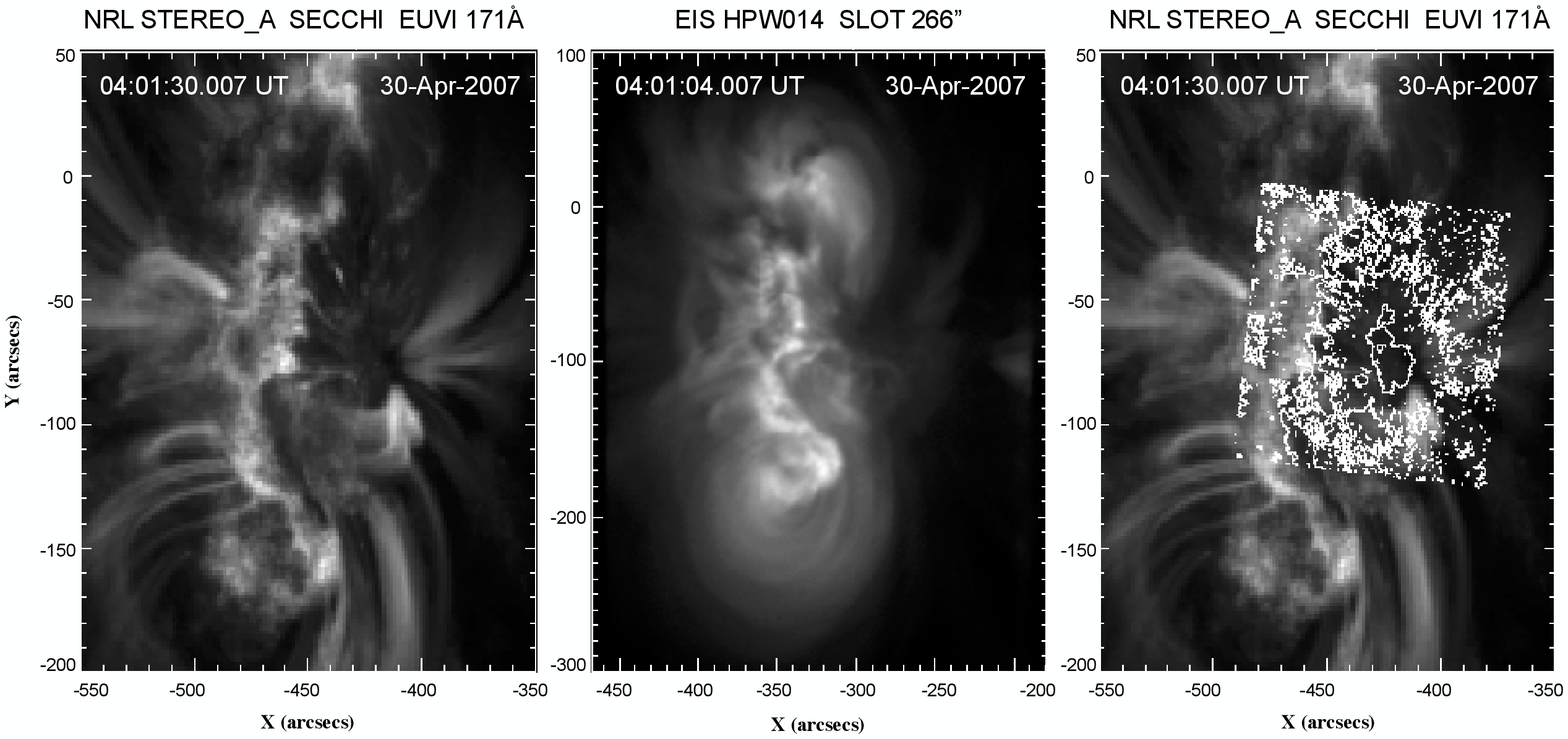}
\caption{EUVI 171 \AA\ image (left); EIS 284 \AA\ slot image (centre) 
and Ca II contours overlaid on EUVI 171 \AA\ (right).The corona above the light-bridge shows 
enhanced intensity
relative to the surrounding umbra. }
\end{figure}

\subsection{EIS velocities}
Supersonic downflows have been reported in the light-bridge on 1 May 2007 by \cite{louis2009}, and we 
note that enhanced (not supersonic) velocities are also seen on 30 Apr 2007 (Fig.1).  \cite{louis2009} 
suggest the supersonic downflows provide evidence of chromospheric reconnection, since a good (not perfect) 
correlation is seen with Ca II jets from the light-bridge. In the EIS data we see that the light- bridge 
is located in regions that show significant outflows; the largest ($\sim$40 km/s) being seen in He II. Figure 
3 shows a series of EIS intensity images from 30 April 2007 on the top row, with temperature of 
formation increasing from left to right (He II, Fe VIII, Fe XII and Fe XV). Beneath each intensity image is the 
corresponding relative velocity map derived from a single component fit. 
We caution that the He II line is a complex blend including contributions from Si X and Fe X on the red-side
 and that we have not fitted the blends. However, the 
dominant blends are to the red, and thus more likely to produce enhanced downflows rather than upflows.

\begin{figure}
\centering
\includegraphics[angle=-90,width=1.1\linewidth]{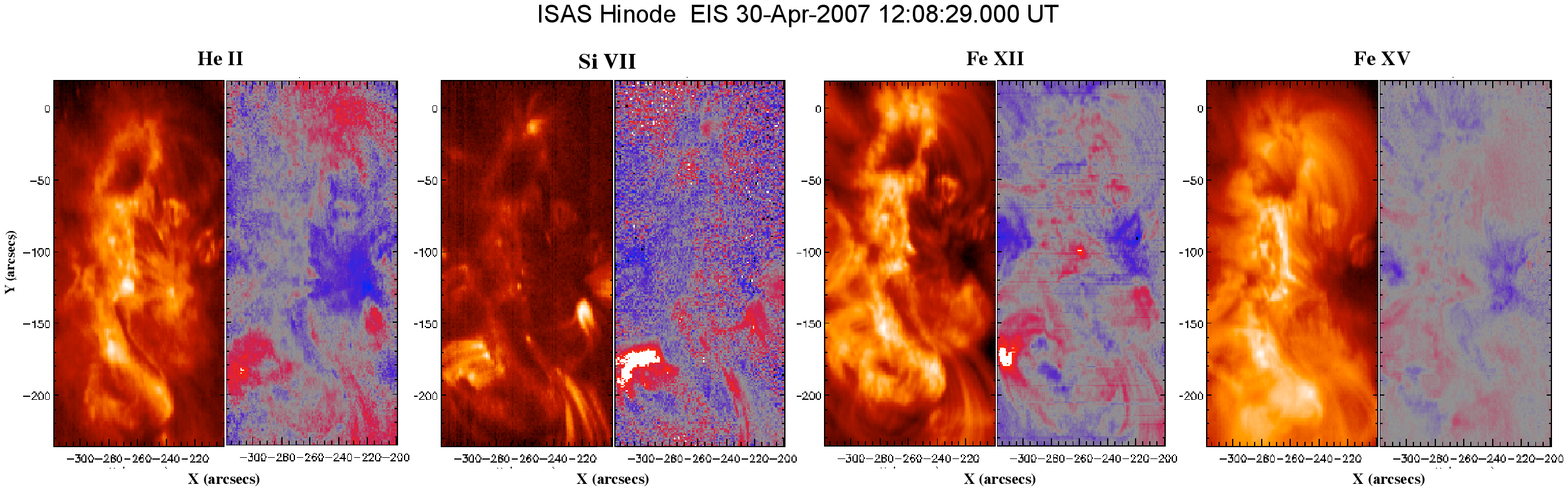}
\caption{Top:EIS raster scans showing intensity of the active region in lines of increasing 
temperature of formation (from left to right: He II, Si VII, Fe XII and Fe XV). The raster began
at 12:08 on 30 April 07 and EIS scans from right to left. Bottom: corresponding Doppler velocity images.}
\end{figure}

\section{Summary \& Conclusions}
We find evidence for increased intensity above the light-bridge in 
STEREO EUVI 171\AA\ data, in EIS He II ( 256 \AA\ ) and Fe XII (195 \AA\ ) rasters, 
and also in the EIS 266$^{\prime\prime}$ Fe XV 284 \AA\ slot images, but not in Si VII 275 
\AA\ raster images.  There are some puzzling inconsistencies, 
but nevertheless indications that enhanced heating above light-bridges 
extends higher than previously thought into the corona. Enhanced upflows are seen with 
EIS in the regions above the light-bridge.These upflows are 
strongest in the He II 256 \AA\ line, and are also located in the vicinity of the outflows that 
have been identified as potential contributors to the slow solar wind \citep[e.g.][and references therein]{baker2009}.
A more detailed presentation of this work will appear in a forthcoming paper.

\vfill\pagebreak
\end{document}